\documentclass[aps,jstat,twocolumn,superscriptaddress,showpacs]{revtex4-1}
\usepackage{graphicx}
\usepackage{psfrag}
\usepackage{ae}
\usepackage{amsmath,amssymb}
\usepackage{float}
\usepackage{bm}
\usepackage{float}

\begin{document}
\date{\today}

\title{Predicting the Final Stationary State of Sheared 2D Euler Fluids}

\author{Renato Pakter}
\author{Yan Levin}
\affiliation{
  Instituto de F\'{\i}sica, UFRGS, 
  Caixa Postal 15051, CEP 91501-970, Porto Alegre, RS, Brazil  }

\begin{abstract}
The Kelvin-Helmholtz (KH) instability is a fundamental signature of the natural world, manifesting in phenomena ranging from atmospheric clouds to the iconic turbulence of Jupiter's Great Red Spot. While the linear dynamics of this instability is well-understood, the ultimate fate of the fluid --- the final stationary state to which it relaxes --- remains a formidable challenge. In this paper, we present a general non-equilibrium statistical mechanics framework that allows us to predict the asymptotic vorticity distribution of a sheared 2D Euler fluid. By constructing a self-consistent ansatz for the solution of the Vlasov equation, we determine the stationary stream function without explicitly solving the time-dependent Euler equations. We demonstrate that the relaxation leads to a population-inverted core, where high-energy states are fully occupied, surrounded by a tenuous halo of vortices in the lower-energy regions of configuration space. This result establishes the core-halo structure as a robust statistical consequence of the system's evolution, offering a universal framework for understanding the saturation of shear-induced turbulence in inviscid 2D fluids. 
\end{abstract}


\maketitle

\section{Introduction}

Boltzmann-Gibbs (BG) statistical mechanics provides a rigorous framework for describing the equilibrium states of systems with short-range interactions. Given an initial state, the final equilibrium configuration can be predicted solely from conserved quantities, circumventing the need to explicitly solve the dynamical equations of motion. Extending this framework beyond its proven validity has been a central challenge for decades~\cite{On49,Ly67,SoRo91,ChSo96,BoVe12}. A seminal attempt was made by Onsager in 1949, who sought to explain the formation of large-scale vortex structures in 2D turbulence by treating the fluid as a system of interacting vortices~\cite{On49}.

Following the Eulerian framework of Helmholtz~\cite{He67} and Kirchhoff~\cite{Ki83}, the incompressible 2D Euler equations are described by the vorticity field $\Gamma({\bf r}, t) = [\nabla \times {\bf u}({\bf r}, t)] \cdot \hat{\bf z}$, which is advected by the flow, 
\begin{equation}
\frac{d\Gamma}{dt} = 0.
\label{advect}
\end{equation}
The incompressibility condition allows for the introduction of a stream function $\psi({\bf r}, t)$, where ${\bf u} = \nabla \times (\psi \hat{\bf z})$. This yields the Poisson equation:
\begin{equation}
\nabla^2 \psi({\bf r}, t) = -\Gamma({\bf r}, t),
\label{poisson}
\end{equation}
the solution of which is obtained via the appropriate Green function:
\begin{equation}
\psi({\bf r}, t) = \int G({\bf r}, {\bf r'}) \Gamma({\bf r'}, t) \, d{\bf r'}.
\label{e2}
\end{equation}
Representing the vorticity as a collection of point vortices, $\Gamma({\bf r}) = \sum \Gamma_i \delta({\bf r} - {\bf r}_i(t))$, reveals that the dynamics is Hamiltonian:
\begin{equation}
\Gamma_i \dot{x}_i = \frac{\partial \mathcal{H}}{\partial y_i} \,\, ; \,\, \Gamma_i \dot{y}_i = -\frac{\partial \mathcal{H}}{\partial x_i},
\label{e3}
\end{equation}
where $\mathcal{H} = \sum_{i<j} \Gamma_i \Gamma_j G(\mathbf{r}_i, \mathbf{r}_j)$ is the Kirchhoff function. In open space, the interaction between vortices is logarithmic, which is extremely long-range.

To predict the final stationary state to which an incompressible fluid confined in a 2d domain of area $A$ will relax, Onsager appealed to the Boltzmann-Gibbs equilibrium statistical mechanics. He observed that, since the vortex ``phase space" is bounded, the number of microstates is finite. The maximum entropy state corresponds to a fully disordered vortex distribution with energy $E_c$. For $E>E_c$ any inhomogeneous vortex distribution will, therefore, have lower entropy than $S(E_c)$. Since 
\begin{equation}
\frac{1}{T} = \frac{\partial S}{\partial E}
\label{temp_deriv}
\end{equation}
this means that such states will have negative absolute temperature! Since in equilibrium, the probability of a given vortex configuration is proportional to the Boltzmann weight, a negative temperature state~\cite{FrWa15} implies clustering of vortices of the same sign, offering a theoretical basis for the spontaneous emergence of coherent structures in 2D turbulence.

Onsager's argument relies on the ergodicity of vortex motion. A fundamental result in fluid dynamics~\cite{MaPu94,EySp06} establishes that smooth solutions of the 2D Euler equation can be approximated over a finite time interval using vorticity 
\begin{equation}
\Gamma({\bf r},t)=\sum \Gamma_i \delta ({\bf r} -{\bf r}_i(t)),
\label{vort_sum}
\end{equation}
with $\Gamma_i \sim \pm 1/N$ and $N \rightarrow \infty$. This corresponds to the thermodynamic limit for systems with long-range interactions~\cite{CaDa09,CaDa14,LePa14}. To predict the final states in this regime, classical statistical frameworks like the Lynden-Bell (LB) and Miller-Robert-Sommeria (MRS) theories~\cite{Ly67,SoRo91,BoVe12} were developed to find global maximum entropy configurations constrained by the Casimir invariants~\cite{Ro05}. Thanks to sophisticated numerical algorithms, it is now possible to compute these strict LB~\cite{TeFa23} and MRS statistical equilibria~\cite{RyIs24,RyIs26} directly from arbitrary initial profiles. However, real long-range interacting systems routinely fail to relax to these idealized maximum entropy states or to traditional Boltzmann-Gibbs equilibrium. Instead, they become trapped in non-ergodic, quasi-stationary states that preserve a strong memory of the initial particle distribution~\cite{CaDa09,CaDa14,LePa14}. Consequently, the true asymptotic behavior of 2D turbulence cannot be captured by the global entropy maximization frameworks, but must instead be described by the non-equilibrium statistical mechanics of long-range interacting systems~\cite{LePa14}. In this paper, we present such a theory to predict the final stationary state of a 2D Euler fluid undergoing a Kelvin-Helmholtz (KH) instability~\cite{Ke71,He68}.  Our motivation is to understand large-scale vortex formation, such as the one observed in Jupiter's Great Red Spot, the origin of which is believed to arise from the Kelvin-Helmholtz instability of the Jovian jets.  
The approach presented in this work is based on the core-halo ansatz, which has been previously shown to accurately describe the relaxation to the final stationary state in long-range interacting systems~\cite{LePa14} ranging from collisionless plasmas~\cite{LePa08}, to self-gravitating clusters~\cite{TeLe10,TeLe11}, and the relaxation of a single perturbed Kirchhoff vortex~\cite{PaLe18}. While our previous vortex study focused on the isolated core-halo dynamics of a pre-existing elliptical patch, the present work addresses the full non-linear evolution—from the initial linear Kelvin-Helmholtz instability and multi-vortex filamentation of a distributed shear layer to its asymptotic stationary state.

\section{Shear Flow and Linear Stability Analysis}

To maintain analytical tractability, we will replace the spherical planetary geometry, by an infinitely long cylinder of radius $R$. This is a reasonable approximation since the width of the equatorial jet streams is much smaller than the Jupiter's radius $R$. The vortex position on the surface of the cylinder can be expressed in terms of the azimuthal and vertical coordinates, $\theta$ and $z$, as ${\bf r}=R \cos\theta\hat {\bf x}+R \sin\theta\hat {\bf y}+z \hat {\bf z}$. The fluid velocity is then ${\bf u}({\bf r}, t)=\nabla \times \psi({\bf r}, t)\hat{\pmb{\rho}}$, where $\hat{\pmb \rho}$ is the unit vector normal to the surface of the cylinder.

\begin{figure}
\begin{center}
\includegraphics[scale=.6]{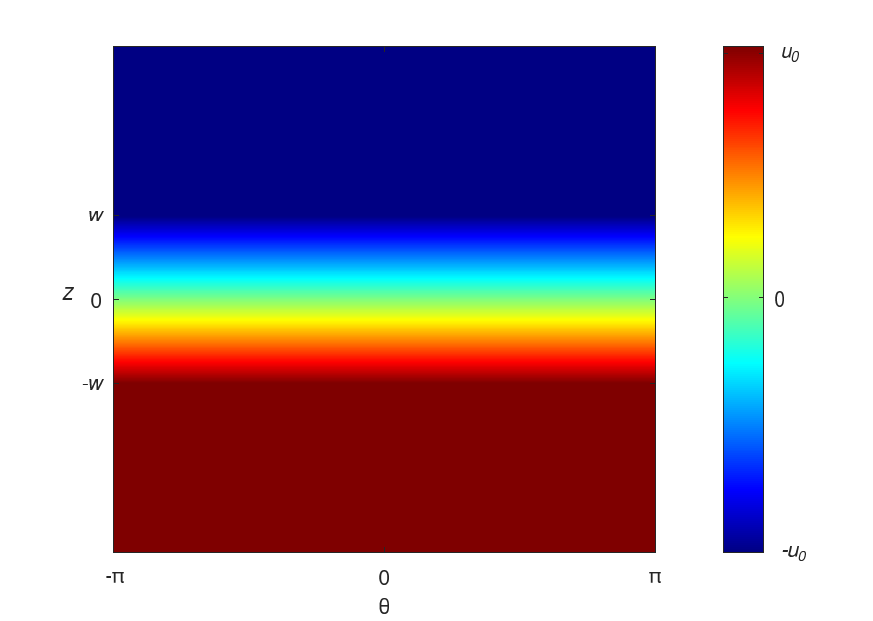}
\caption{
Initial velocity field of a sheared 2D fluid described by Eq. (\ref{e_vel}).  Color map provides the azimuthal velocity.  The shear zone is located between $-w< z<w$.  For $z>w$  velocity is $-u_0$ and below $z<-w$ it is $+u_0$.  The coordinate $z$ is measured in units of $R$.}
\label{shear}
\end{center}
\end{figure}

To characterize the onset of turbulence, we first define the unperturbed shear layer of width $2w$ on a cylindrical surface, see Fig. (\ref{shear}). The velocity field for this layer is given by:
\begin{equation}
 {\bf u}_0({\bf r})=-\hat{\pmb \theta}\left[{z u_0\over w} \,\Theta (w^2-z^2)+u_0 \, {\rm sign}(z)\Theta (z^2-w^2)\right].
\label{e_vel}
\end{equation}
This flow corresponds to a ``water-bag" vorticity distribution $\Gamma_0 ({\bf r})=\gamma_0\Theta [w^2-z^2]$, with $\gamma_0=u_0/w$, generated by the stream function:
\begin{equation}
 \psi_0({\bf r})=-{(w^2+z^2) u_0 \Theta (w^2-z^2)+2w|z| u_0\Theta (z^2-w^2)\over 2 w}.
\end{equation}
The total vortex strength is $4 \pi R u_0$. The kernel for the Poisson equation in this cylindrical geometry is the Green function~\cite{PaLe21}:
\begin{equation}
 G({\bf r},{\bf r}')=-{1\over 4\pi}\log \left [2 \cosh\left ({z-z'\over R}\right )-2 \cos(\theta-\theta ')\right ].
\end{equation}

The KH linear instability is usually studied using a perturbative solution of 2d Euler equations~\cite{Re80,KuCo08}. Here however, we will use the approach introduced in our earlier work to determine the stability limit of the shear flow directly in terms of the vortex distribution~\cite{PaLe21}.  We consider small perturbations of vorticity distribution upper and lower boundaries --- ($u, l$), respectively --- $z^u(\theta,t)=w+{\cal F}^u(\theta,t)$ and $z^l(\theta,t)=-w+{\cal F}^l(\theta,t)$. Expanding the upper and lower borders perturbations in Fourier series and linearizing the governing Euler equations results in a closed system of equations for the perturbation amplitudes of a given Fourier mode, $\mathbf{C}_n=(a_n^l, b_n^l, a_n^u, b_n^u)^T$, respectively:
\begin{equation}
\frac{d{\bf C}_n}{dt}={\cal M}_n \cdot {\bf C}_n.
\end{equation}
The stability matrix ${\cal M}_n$ is found to have eigenvalues given by
\begin{equation}
\gamma_n = \pm \frac{\eta_0}{2}\sqrt{\beta_n^2-\alpha_n^2},
\end{equation}
where the coefficients are defined as
\begin{equation}
\alpha_n = 1 - \frac{2nw}{R}, \quad \beta_n = e^{-\frac{2nw}{R}}.
\end{equation}
The stability is lost when $\beta_n^2 > \alpha_n^2$. The first mode ($n=1$) becomes unstable at $w/R \approx 0.64$. This threshold defines the regime where the initial shear flow will become unstable and relax to a non-equilibrium stationary state which will be analyzed below.

\section{Molecular Dynamics Simulations and Vlasov Relaxation}

To explore the validity of the theory, we performed molecular dynamics (MD) simulations using a Vortex-in-Cell (VIC) algorithm~\cite{vic,ChZa73,DrSc08,DrSc09}. The algorithm simulates the evolution of the vorticity field. The computational domain is discretized using a $250\times250$ grid and $5\times10^5$ particles. At each time step, the vorticity carried by each particle is distributed among the four nearest grid points using bilinear weighting. The Poisson equation for the stream function is then solved on the grid using a successive-over-relaxation (SOR) method. The velocity at the particle positions is obtained from the stream function using a nine-point polynomial interpolation scheme.
The particles are advanced using an adaptive time step based on an embedded fourth- and fifth-order Runge-Kutta method. The stream function is updated {\it continuously} during the Runge-Kutta integration, rather than being held fixed over an entire time step. This procedure provides accurate particle trajectories while maintaining good conservation properties. As a numerical accuracy check, the maximum relative variation of the total energy throughout the simulations was approximately 0.01\%.  

\begin{figure}
\begin{center}
\includegraphics[scale=.45]{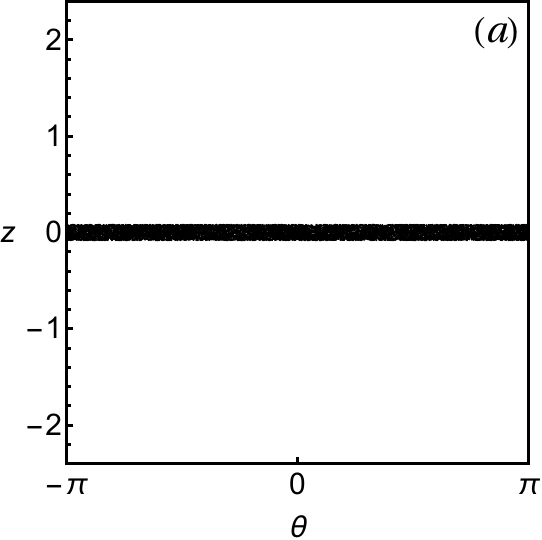}
\includegraphics[scale=.45]{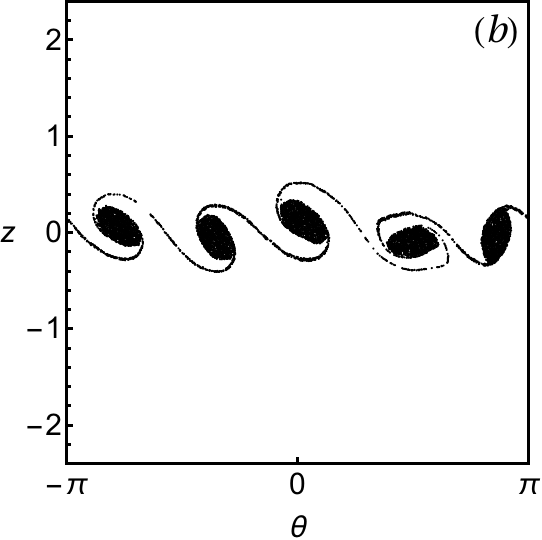}
\includegraphics[scale=.45]{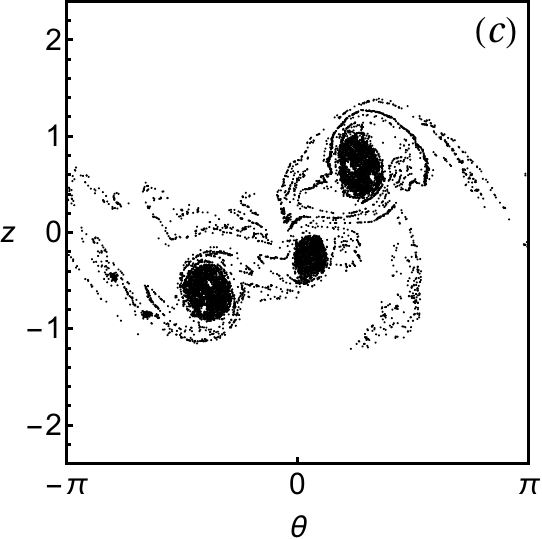}
\includegraphics[scale=.45]{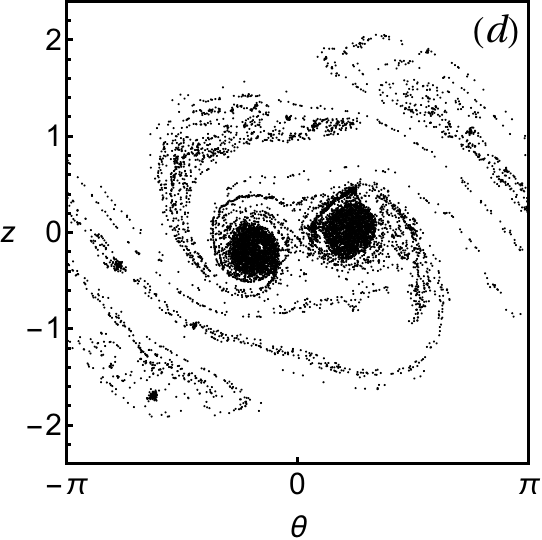}
\includegraphics[scale=.45]{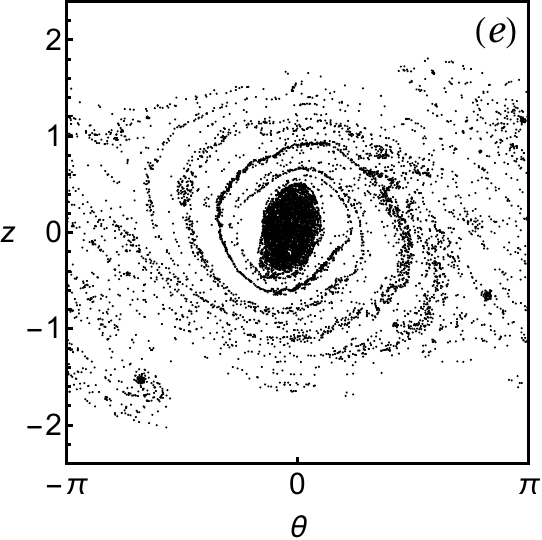}
\includegraphics[scale=.45]{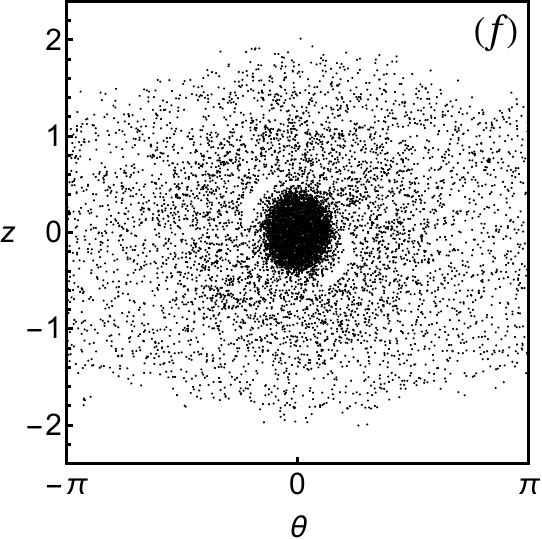}
\caption{Temporal evolution following KH instability for initial water-bag vortex distribution with $w/R=0.08$.  For this $w$, the theory predicts that the most unstable mode is $n=5$.  Snapshots of the phase space at:  $t=0$ (a); $t=6.0$ (b); $t=18.0$ (c); $t=35.0$ (d); $t=48.0$ (e); and $t=900.0$ (f). The coordinate $z$ is in units of $R$.  The final stationary state has a core-halo form.}
\label{snaps}
\end{center}
\end{figure}

Fig.~\ref{snaps} shows the temporal evolution of a distribution with $w/R=0.08$. For this $w$ the theory predicts that mode $n=5$ is most unstable, i.e. has largest $\gamma_n$. Indeed, the simulations show that initial instability results in $5$ macroscopic vortices, which then merge through the process of filamentation [panels (c) and (d)], eventually relaxing into a stationary state characterized by a dense coherent vortex core surrounded by a tenuous halo [panels (e) and (f)].

In the thermodynamic limit, $N \rightarrow \infty$, $\Gamma_v =\frac{4 \pi R u_0}{N} \rightarrow 0 $, 
the relaxation of the initial water-bag vortex distribution is governed {\it exactly} by the Vlasov equation~\cite{BrHe77},
\begin{equation}
 \frac{\partial f}{\partial t}-\frac{1}{R}\frac{\partial \psi}{\partial \theta} \frac{\partial f}{\partial z} +
\frac{1}{R} \frac{\partial \psi}{\partial z} \frac{\partial f}{\partial \theta}=0 \,.
\end{equation}
This equation is identical to the condition that vortices are advected by the flow, so that the vortex gas evolves as an incompressible fluid. The incompressibility constraint of the Vlasov dynamics implies that the distribution function can never exceed the initial vortex density $\eta_0=N/(4 \pi R w)$. The relaxation to equilibrium comes through the process of Landau damping~\cite{La46, Gl94} in which some of the vortices are captured by the parametric resonances and are ejected from the main cluster. This is similar to the process of collisionless relaxation observed across a wide variety of systems with long-range interactions~\cite{Ch22} such as gravitational clusters~\cite{TeLe10,TeLe11,JoWo11}, magnetically confined plasmas~\cite{LePa08}, spin systems~\cite{PaLe11,Ro16} etc. The difference in the case of the vortices is that the parametric resonances, and the associated chaotic diffusion, eject vortices into the low energy regions of the phase space, instead of the high energy regions for the usual many body long-range interacting systems. This results in the process of ``evaporative heating" of the core~\cite{SuDa14,GrDa18,PaLe18}.  

\begin{figure}
\begin{center}
\includegraphics[scale=.7]{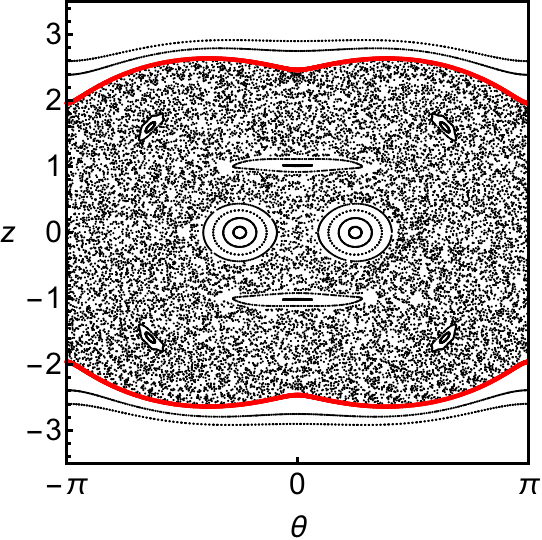}
\caption{
The Poincar\'e map of non-interacting test point vortices subject to the potential produced by two macroscopic point vortices each of strength $2 \pi R u_0$, evolving under the Hamilonian dynamics, Eq. (\ref{e3}).  Initial separation between the two interacting vortices is $\Delta\theta=\pi/2$. The chaotic diffusion of test-vortices is confined by the invariant KAM surface represented by the thick red curve, whose maximum vertical displacement is $z_{max} \approx 2.6 R$. The coordinate $z$ is measured in units of $R$.}
\label{poinc}
\end{center}
\end{figure}

To understand better the process of halo formation, we study the dynamics of non-interacting test vortices. We first observe that the ejected vortices move farthest in the final stages of the relaxation process, when the large scale vortex structures of the original KH instability have fused together leaving only two large cores, see panel (d) of Fig. \ref{snaps}. We will, therefore, study the dynamics of {\it non-interacting} test particles moving in a potential produced by two {\it interacting} macroscopic point vortices of strength $\Gamma_m=2 \pi R u_0$, separated by a distance $R \Delta \theta$. Solving the equations of motion, Eq.(\ref{e3}), we find that the two macroscopic vortices rotate around their center of mass with period $T$, producing a periodic potential felt by the test vortices. The Poincar\'e plot of the test vortex dynamics --- stroboscopic plot in which positions of the test vorticis are plotted every $T/2$ --- has a very complicated chaotic structure which, however, is bounded by a Kolmogorov-Arnold-Moser (KAM) curve~\cite{LiLi92} designated by a thick red solid line in Fig. \ref{poinc}. We find that the maximum $z$-distance that a test vortex can achieve starting from initial water-bag distribution is $1.8 R<z_{max}<2.6 R$, depending on the separation between the two rotating macro-vortices. This is consistent with the full MD simulations of $N$ interacting vortices. The maximum $z$-distance reached by any vortex in MD simulations is observed to be $z_{max} \approx 2.1 R$ for all KH unstable shear flows.

\section{Core-Halo Ansatz and Self-Consistent Numerical Solution}

The mechanism of the core-halo formation is now evident. The parametric resonances eject vortices into the low energy phase space regions, far from the core. To compensate for this, other vortices must move into the high energy core. However, because of the incompressibility of the vortex gas the density of the core can't exceed that of the initial vortex distribution $\eta_0=N/4 \pi R w$. The process of evaporative heating will continue until all the high energy states, up to the ``Fermi energy", are occupied up to the maximum density $\eta_0$. Based on these insights we can now propose an ansatz core-halo solution to the stationary Vlasov equation:
\begin{equation}
f_{ch}({\bf r})=
\begin{cases}
\eta_0,\, &{\rm for}\, \psi_{ch}({\bf r})\ge \epsilon_F\\
\chi \left[ \psi_{ch}({\bf r})-\epsilon_h \right] ,\, &{\rm for}\, \epsilon_h < \psi_{ch}({\bf r})< \epsilon_F\\
0 ,\, &{\rm for}\,  \psi_{ch}({\bf r}) < \epsilon_h
\end{cases}
\label{fch}
\end{equation}
where $\psi_{ch}({\bf r})$ is the self-consistent stream function obtained by substituting $\Gamma_{ch}({\bf r})=\Gamma_v f_{ch}({\bf r})$ into the Poisson equation (\ref{poisson}). $\epsilon_F$ is the Fermi energy that determines the extent of the core, and $\epsilon_h$ is the halo energy obtained from the location of the KAM surface in the Poincar\'e plot of test particle dynamics, expressed as
\begin{equation}
\epsilon_h = \psi_{ch}(\theta=0, z=z_{max}).
\end{equation}
The parameter $\chi$ is associated with the halo vortex density, which is assumed to have a linear dependence on $\psi_{ch}({\bf r})$ and to vanish at $\psi_{ch}({\bf r})=\epsilon_h$. It is important to note that the core-halo distribution depends on $\theta$ and $z$ only through the one particle energy --- the stream function, which is time independent in the stationary state. Therefore, $f_{ch}({\bf r})$ is a stationary solution of the Vlasov equation, $df_{ch}({\bf r})/dt=0$. The values of $\epsilon_F$ and $\chi$ are yet unknown. However, they can be determined by imposing the conservation of the total vorticity and the energy, namely,
\begin{equation}
\int f_{ch} ({\bf r})d^2{\bf r}=N,
\label{nch}
\end{equation}
\begin{equation}
{\Gamma_v\over 2} \int f_{ch}({\bf r}) \psi_{ch} ({\bf r})d^2{\bf r}={\cal E}_0,
 \label{ech}
\end{equation}
where the initial energy is given by
\begin{equation}
{\cal E}_0 = \frac{1}{2} \int \Gamma_0({\bf r}) \psi_0({\bf r}) d^2{\bf r}.
\end{equation}
To determine the core-halo stationary state we must self-consistently solve the Poisson equation  for $\psi_{ch}({\bf r})$ obtained when Eq.~(\ref{fch}) is substituted into Eq.~(\ref{poisson}), with the values of $\epsilon_F$ and $\chi$ satisfying Eqs.~(\ref{nch}) and (\ref{ech}).  The stream function is taken to vanish at the walls located at $\pm z_w$. The precise position of the wall does not influence the resulting distribution as long as  $|z_w|\gg 1$.    To this end we have developed a code that solves the Poisson equation on a 2d grid using successive over-relaxation, starting from an initial guess for the distribution function of the form Eq. (\ref{fch}). At each SOR iteration we update the vortex distribution according to Eq.~(\ref{fch}), while $\epsilon_F$ and $\chi$ are updated according to
\begin{align}
\epsilon_F^{m+1} &= \epsilon_F^{m} \left[ 1 - \alpha \left( 1 - \frac{{\cal E}_{ch}^m}{{\cal E}_0} \right) \right], \\
\chi^{m+1} &= \chi^{m} \left[ 1 + \alpha \left( 1 - \frac{N_{ch}^m}{N} \right) \right],
\end{align}
where ${\cal E}_{ch}^m$ and $N_{ch}^m$ are the energy and the number of vortices calculated using $f_{ch}$ at iteration $m$. For a sufficiently small value of $\alpha$, the method converges to a fully self-consistent solution of the problem: as $m \rightarrow \infty$, ${\cal E}_{ch}^m \rightarrow {\cal E}_{0}$ and $N_{ch}^m \rightarrow N$. In practice convergence is very fast.

\begin{figure}
\begin{center}
\includegraphics[scale=.4]{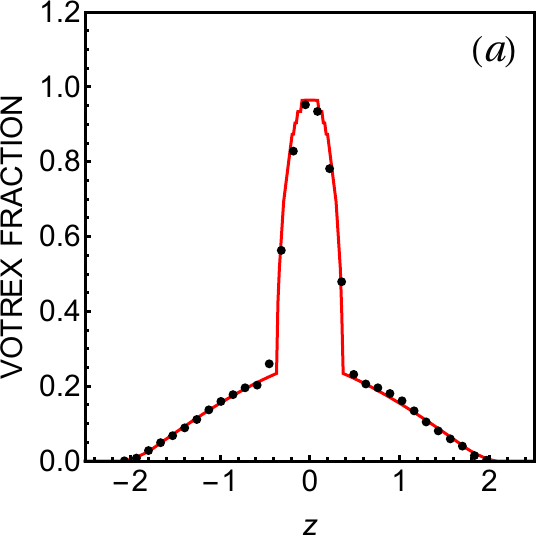}
\includegraphics[scale=.4]{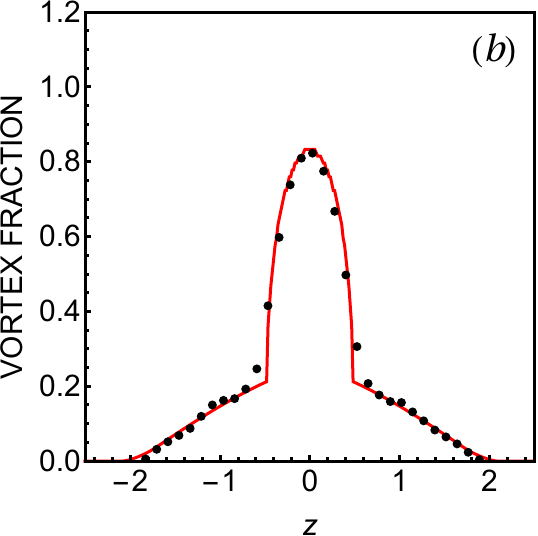}
\caption{Comparison between the stationary transverse vortex fraction $n(z)$ obtained using the core-halo distribution (red solid curve) and the molecular dynamics simulations (black circles) for  (a) $w/R=0.08$ (most unstable mode $n=5$); and (b) $w/R=0.13$ (most unstable mode $n=3$). The numerical results are obtained by averaging over 10 different times after the system has reached a quasi-stationary state. The coordinate $z$ is measured in units of $R$.}
\label{corehalo}
\end{center}
\end{figure}

In  Fig. \ref{corehalo} we compare the results obtained from MD simulations with those predicted by the theory. To perform a quantitative comparison, we define the transverse vortex fraction as
\begin{equation}
n(z) = \frac{R}{N} \int f_{ch}({\bf r}) d\theta,
\end{equation}
corresponding to the fraction of vortices located between $[z,z+dz]$. A very good agreement with simulations is found. 
It is interesting to note that while for other long-range interacting systems such as magnetically confined plasmas, gravitational clusters and Hamiltonian mean-field model, the halo distribution function is constant between Fermi and halo energies~\cite{LePa14}, in the case of vortices resulting from KH instability the halo distribution is linear in stream function, see  Eq. (\ref{fch}). This form is universal, independent of the width of the shear zone, and of which Fourier mode is most unstable, see Fig. \ref{corehalo}. 

\begin{figure} [H]
\begin{center}
\includegraphics[scale=.5]{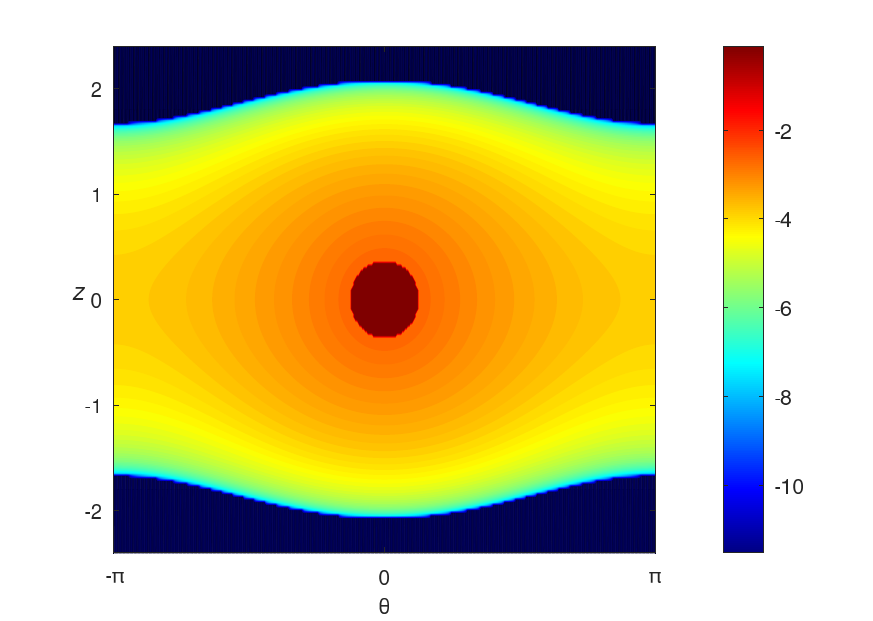}
\includegraphics[scale=.35]{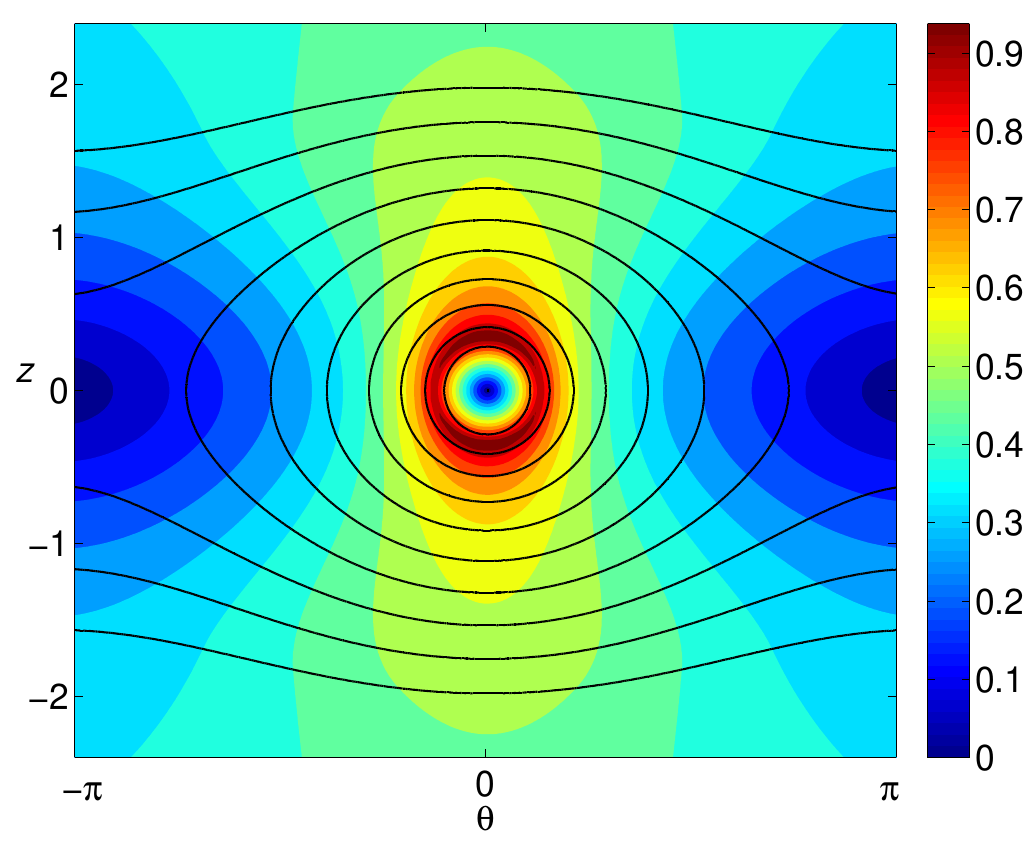}
\includegraphics[scale=.4]{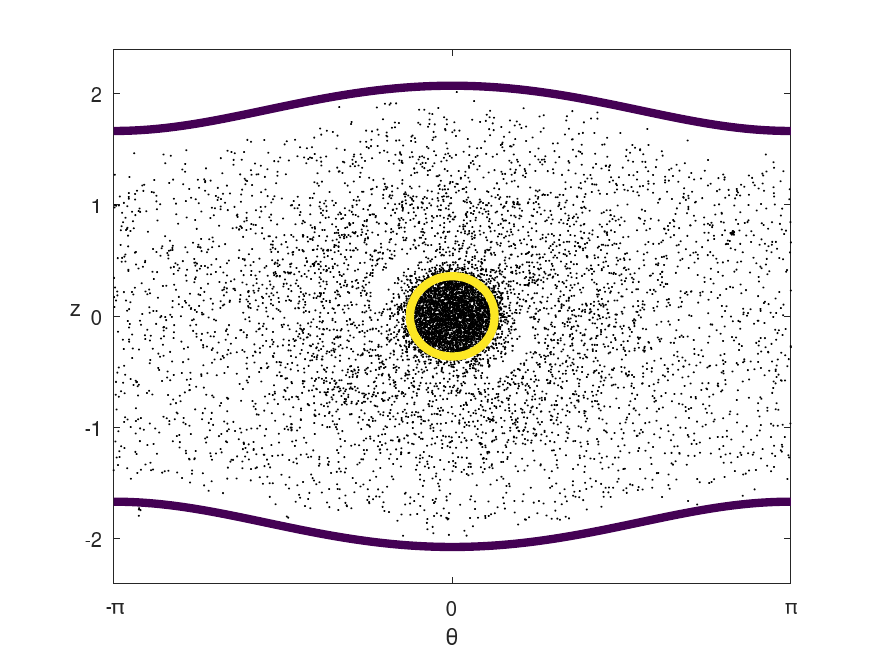}
\caption{The stationary state resulting from unstable shear flow described by Eq. (\ref{e_vel}) with $w/R=0.08$.  The coordinate $z$ is measured in units of $R$. Top panel is the vorticity field, $\Gamma({\bf r})$. Color map in log scale indicates the normalized strength of the field.   Middle panel shows the fluid velocity calculated using the stream function $\psi_{ch}$ obtained from the core-halo distribution Eq. (\ref{fch}).  The black solid curves are the stream lines, equipotentials of $\psi_{ch}$. The color map indicates a normalized speed of the fluid at a given position. We note that the speed vanishes at the center of the core -- eye of the hurricane -- and at the point opposite to it ($\theta=\pm\pi$) and reaches its maximum at the core boundary.  Bottom panel explicitly compares the boundaries for the core (yellow solid curve) and of the halo (purple solid curves) predicted by the theory to the vertex distribution obtained using VIC simulations. }
\label{final}
\end{center}
\end{figure}

A precise knowledge of the stream function allows us to easily calculate the fluid velocity in the final stationary state. In top panel of Fig. \ref{final} we show the vorticity field and in the bottom panel the stream lines -- equipotentials of $\psi_{ch}$ -- with the color code indicating the fluid speed. Note a characteristic formation of the ``eye of the hurricane" in the center of the core. 

\section{Conclusion and Outlook}

We have presented a non-equilibrium statistical mechanics framework that provides a predictive theory for the asymptotic stationary state of a 2D Euler fluid following the KH instability. By mapping the long-term vortex relaxation to a self-consistent Vlasov-Poisson problem, we have demonstrated that the system consistently relaxes to a universal core-halo architecture. We emphasize that this framework describes the inviscid, coarse-grained quasistationary states resulting from collisionless relaxation. The robustness of this state, independent of the initial instability mode $n$, suggests that the core-halo distribution serves as a global attractor for these ideal 2D fluid dynamics.

This framework may provide a new insight into the origin of coherent structures like Jupiter's Great Red Spot as well as offers a new perspective for laboratory-scale 2D turbulence.  Future extensions of this model, particularly to non-cylindrical geometries or systems with modified confinement potentials, will be critical for testing the universality of this relaxation process. Furthermore, this approach can be readily generalized to more complex initial conditions, such as jet flows composed of adjacent, counter-rotating water-bag vorticity layers. In such cases, preliminary analysis suggests that the fluid relaxes to a stationary state -- viewed in the co-moving frame -- characterized by the coherent motion of macroscopic vortices of opposite sign. We anticipate that the self-consistent framework presented here can be adapted to derive these stationary stream functions, a task we reserve for future work.

\acknowledgments
This work was partially supported by the CNPq.

\end{document}